\documentclass[aps,prb,twocolumn,showpacs,superscriptaddress]{revtex4-1}  
\usepackage{amsmath}
\usepackage{amssymb}
\usepackage{graphicx}
\usepackage{epstopdf}
\usepackage{suffix}
\usepackage{mathtools}

\begin{document}

\title{Chemical potential of an antiferromagnetic magnon gas}

\author{Benedetta Flebus}
\address{Department of Physics and Astronomy, University of California, Los Angeles, California 90095, USA}

\begin{abstract}  

Understanding the statistics of quasi-particle excitations in magnetic systems is essential for exploring new magnetic phases and collective quantum phenomena. While the chemical potential of a ferromagnetic gas has been extensively investigated both theoretically and experimentally,  its antiferromagnetic counterpart remains uncharted.  Here, we derive the statistics of a two-component $U$(1)-symmetric Bose gas and apply our results to an axially-symmetric antiferromagnetic insulator. We find that the two magnon eigenmodes of the system are described by an equal and opposite chemical potential, in analogy with a particle-antiparticle pair.  Furthermore, we derive the thermomagnonic torques describing the interaction between the coherent and incoherent antiferromagnetic spin dynamics. Our results show that the magnitude and sign of the chemical potential can be tuned via an AC magnetic field driving resonantly one of the magnon modes. Finally, we propose NV-center relaxometry as a method to experimentally  test our predictions. 
\end{abstract}


\maketitle

%
%\begin{figure}[b!]
%\includegraphics[width=1\linewidth]{fig11}
%\caption{\textbf{Single-wall carbon nanotube  on top of a graphene layer.} The nanostructures are aligned in an AB-stacked arrangement: the   atoms located at the B sites of the carbon nanotubes are above the A sites of the lower graphene layer. Top left:  lattice structure of an unrolled carbon nanotube. The nanotube  lies on top of the graphene layer at an angle $\eta$ and its geometry is specified by a primitive translation vector $\vec{T}$ along the nanotube axis, and the chiral vector $\vec{L}$  perpendicular to  $\vec{T}$.  Bottom left: $\vec{\mathcal{G}}_{2}$ and $\vec{\mathcal{G}}_{3}$ are the reciprocal lattice vectors connecting a Dirac point to the two other equivalent ones.}
%\label{Fig1}
%\end{figure} 

\section{Introduction}

In the last decade, magnetic insulators have been surfacing as a promising platform for long-range information transport and emergent collective quantum phenomena~[\onlinecite{review}]. 
While electronic motion is prohibited in these systems, the electron's spin can  be transported by magnons, i.e., quasi-particles corresponding to the quantized fluctuations of the magnetic order parameter. Long-range magnon transport has been observed in yttrium iron garnet (YIG) that was driven either thermally~[\onlinecite{Cornelissen2015}, \onlinecite{Giles2015}] or electrically~[\onlinecite{Cornelissen2015}, \onlinecite{Gross2015}].
Prior these measurements, magnons have been described as a Bose gas with no chemical potential, due to the non-spin-conserving magnon-phonon scattering that occurs in all magnetic systems. Recently, it has been instead established that the observed long-range magnon transport can not be explained without invoking a magnon chemical potential~[\onlinecite{Cornelissen2016}]. This quantity parametrizes a long-living non-equilibrium magnon state that can be realized when the non-spin-conserving interactions are much weaker than the spin-conserving scattering processes that drive the thermalization of the magnon gas to a state described by Bose-Einstein statistics.

The first direct measurement of the magnon chemical potential has been performed by C. Du \textit{et al}~[\onlinecite{Du2017}]. Their experimental scheme relies on the interplay between the coherent and incoherent ferromagnetic spin dynamics~[\onlinecite{Flebus2016}]. An AC magnetic field drives the long-wavelength spin dynamics of YIG, which, in turn,  pumps spin angular momentum into the thermal magnon gas, leading to an increase of the  chemical potential  that can be  probed via NV-center relaxometry~[\onlinecite{Casola}].

The introduction of a magnon chemical potential has sparked numerous efforts to realize magnon Bose-Einstein condensation (BEC). Macroscopic occupation of the lowest-energy  mode of a ferromagnetic gas has been achieved via parametric pumping~[\onlinecite{Demokritov}] or microwave drive~[\onlinecite{Du2017}].  
On the other side, \textit{U}(1)-symmetric magnetic insulating systems have been proposed as a host of magnon BEC at equilibrium. This phase has been predicted to support a spin superfluid, i.e., a long-range propagating Goldstone mode emerging from the spontaneous  breaking of \textit{U}(1) symmetry~[\onlinecite{Sonin}].   The lack of dipole-dipole interactions make axially-symmetric antiferromagnets (AFMs)  ideal candidates for the realization of magnon BEC and spin superfluidity~[\onlinecite{Takei,Yuan,Klaui}].  However, despite their promise and abundance in nature~[\onlinecite{Baltz}], these systems have not been subjected yet to the same  scrutiny of their ferromagnetic counterpart. 
\begin{figure}[b!]
\includegraphics[width=0.8\linewidth]{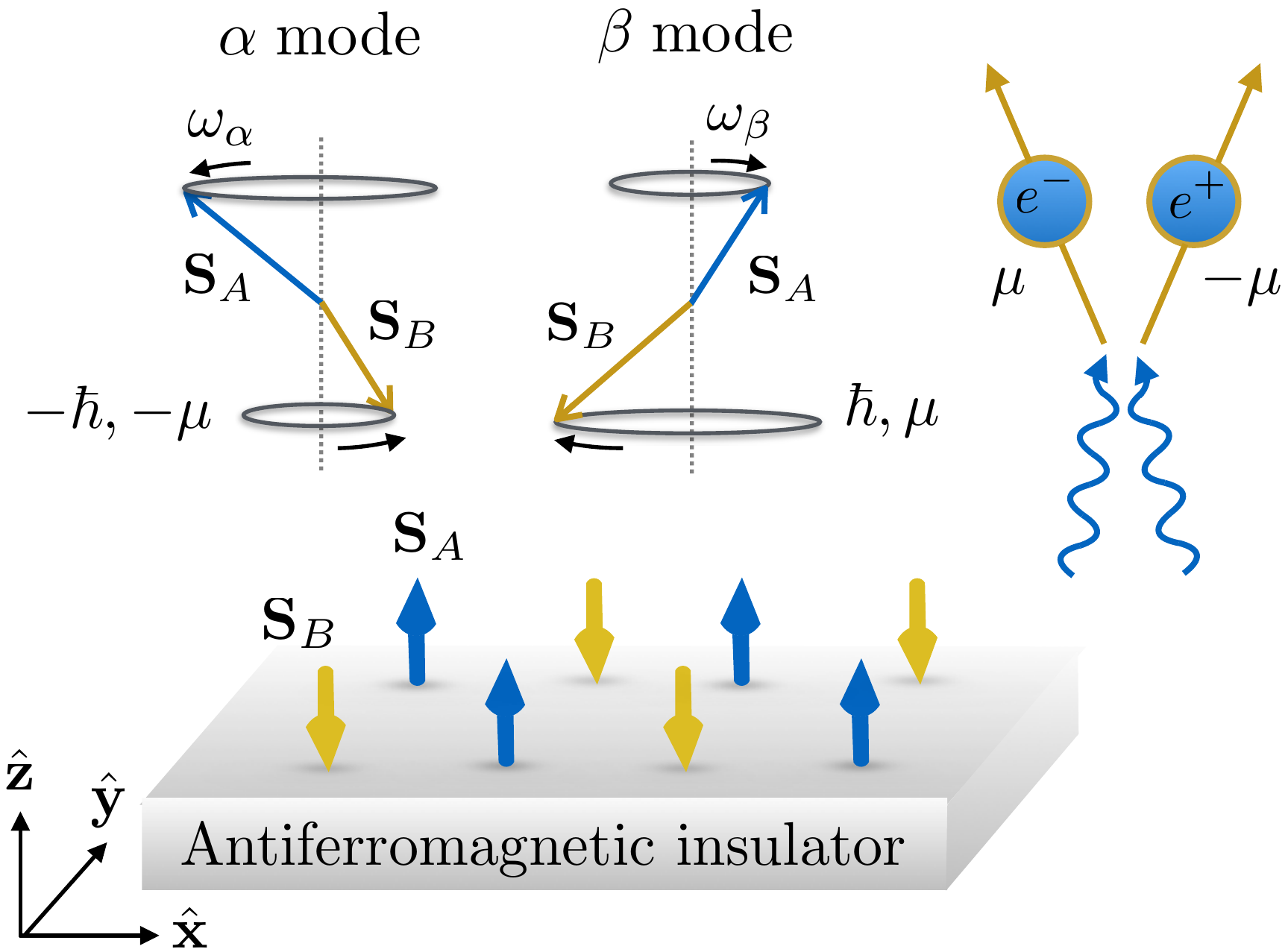}
\caption{$U$(1)-symmetric  antiferromagnetic insulator. The bipartite antiferromagnet is composed of two sublattices A and B, with, respectively, spin  $\mathbf{S}_{A} \parallel \hat{\mathbf{z}}$ and $\mathbf{S}_{B} \parallel -\hat{\mathbf{z}}$. The corresponding Hamiltonian can be diagonalized in terms of two magnon eigenmodes, labelled as  $\alpha$ and $\beta$. The antiferromagnetic mode $\alpha$ ($\beta$), which carries $\mp \hbar$ spin angular momentum, can be visualized as a combination of both up and down spins, precessing counterclockwise (clockwise) at the frequency $\omega_{\alpha}$ ($\omega_{\beta}$). The two antiferromagnetic modes are associated with an equal and opposite chemical potential $\mu$, in analogy with the particle-antiparticle pair illustrated on the left side. }
\label{Fig1}
\end{figure} 

In this work,  we  establish a rigorous definition of the  chemical potential of a $U$(1)-symmetric antiferromagnetic system and we propose an experimental set up for both controlling and probing it.  
The collective excitations of the AF system can be described in terms of two magnon modes carrying opposite spin angular momentum.   In Sec.~II,  we show that the statistics of the antiferromagnetic magnon modes mimics the one of a particle-antiparticle couple, i.e.,  they can be described by an equal and opposite chemical potential, as depicted in Fig.~\ref{Fig1}.  
In Sec.~III, we derive a formalism describing the interplay between coherent and incoherent antiferromagnetic spin dynamics. In analogy with the ferromagnetic case~[\onlinecite{Flebus2016}], we find that the thermomagnonic torques exerted by the thermal magnon cloud and the back-action of the  order parameter  resemble the well-known mechanisms of spin transfer~[\onlinecite{Berger1996}] and spin pumping torques~[\onlinecite{Bauer2002}], respectively. Semiclassically, the dynamics of the two magnon eigenmodes can be described by two order parameters precessing with opposite handedness. A circularly-polarized AC magnetic field can resonantly drive the magnon mode  that precess with the same handedness, and, thus,  control the magnon chemical potential via spin pumping of the coherent spin dynamics into the thermal magnon cloud.  
Based on these findings, we propose a scheme for testing our predictions via NV-center relaxometry in Sec.~IV. Finally, in Sec.~V we present our conclusions and an outlook.

\section{Statistics and model}

In this section, we address the statistics of a two-component \textit{U}(1)-symmetric Bose gas. As a concrete application of our result,  we consider a model Hamiltonian for a bipartite antiferromagnet with uniaxial anisotropy and we derive the distribution functions of the two magnon eigenmodes. Finally, we introduce the Landau-Lifshitz-Gilbert equations corresponding to our Hamiltonian model and discuss the effect of circularly-polarized AC magnetic fields with opposite handedness on the long-wavelength antiferromagnetic magnetization dynamics.

\subsection{Statistics of a two-component \textit{U}(1)-symmetric Bose gas}

Throughout this work, we focus on a  $U$(1)-symmetric antiferromagnetic insulating system. At temperature $T \ll T_{N}$, with $T_{N}$ being the N\'eel temperature, the corresponding  Hamiltonian can be diagonalized in terms of two magnon modes, carrying opposite spin angular momentum. Here, we address the statistics of the two magnon modes in the limit of  weak non-spin-conserving phonon-magnon scattering processes.  Within this assumption, we can treat  the energy, $U$, and the $z$-component of the total spin, $S_{z}$,  of the magnetic system as conserved. 
The total number of available microstates in the bosonic subsystems, labelled as $1$ and $2$, respectively, can be written as 
\begin{align}
\Omega_{1 (2)}=\prod_{n}\frac{(n_{n}+g_{n})!}{n_{n}! (g_{n}-1)!}\,, 
\end{align}
with $n_{n}$  being the particle number in the energy level $\epsilon_{n}$ with degeneracy $g_{n} \gg 1$. The number of microstates of the whole system, i.e.,  $\Omega=\Omega_{1} \Omega_{2}$,
has to be maximized subject to the following constraints 
\begin{align}
\hbar \sum_{i} n_{i} s_{i} +\hbar \sum_{j} n_{j} s_{j} = S_{z}\,, \nonumber \\
\sum_{i} n_{i} \epsilon_{i} + \sum_{j} n_{j} \epsilon_{j} = U\,,
\end{align}
where $s_{i(j)}= \pm 1$  is the spin angular momentum (in units of $\hbar$) carried by the magnons in the subsystem 1(2). For $n_{i (j)} \gg 1$,  we obtain, after averaging over the $g_{i(j)}$ levels in each clump,
\begin{align}
n_{i(j)}=\frac{1}{e^{\beta \epsilon_{i(j)}- \lambda s_{i(j)}}-1}\,, 
\end{align}
where $\lambda$ is a Lagrange multiplier and $\beta=1/(k_{B}T)$, with  $k_{B}$ being the Boltzmann constant. The entropy variation of the  subsystem 1(2) can be derived as
\begin{align}
d \mathcal{S}_{1 (2)}&= k_{B} \sum_{i (j)} \text{ln} \frac{g_{i (j)}+n_{i (j)}}{n_{i (j)}} dn_{i (j)}\,, \nonumber \\
&=-k_{B} \lambda s_{1 (2)} \; dN_{1 (2)}+dU_{1 (2)}-dG_{1 (2)}\,,
\label{84}
\end{align}
where $N_{1(2)}=\sum_{i(j)} n_{i(j)}$  and $G_{1(2)}=\sum_{i(j)} \epsilon_{i(j)}$ are, respectively, the total number of magnons and the Gibbs free energy of the subsystem 1(2). Using Eq.~(\ref{84}), we can identify $\mu_{1 (2)}=\lambda s_{1 (2)} k_{B} T$. Thus, we can assign the chemical potential $\mu_{1(2)}=\pm \mu$ to the magnon mode carrying  spin angular momentum  $\pm \hbar$. By rewriting the entropy variation of the whole system,  $d\mathcal{S}$, we can identify the chemical potential $\mu$ as the energy cost associated with an imbalance between the population of the two magnon species, i.e., 
\begin{align}
\mu= \left[ \frac{\partial U}{\partial (N_{1}-N_{2})} \right]_{\mathcal{S},V}\,,
\label{91}
\end{align}
with $V$ being the system volume.

%When spin non-conserving interactions are accounted for, the system can reach a point where its free energy is minimized with respect to the quasi-particle number. In this case, while the overall chemical potential of the system vanishes, the two magnon modes can be still described by   Equation~(\ref{91}) becomes
%\begin{align}
%\frac{1}{\hbar} \left[ \frac{\partial U}{\partial (N_{1}-N_{2})} \right]_{\mathcal{S},V}=\mu_{1}+\mu_{2}=0\,.
%\end{align} 

\subsection{Hamiltonian model}

For an explicit model of a $U$(1)-symmetric antiferromagnetic system, we consider a bipartite antiferromagnetic  insulator with spins $\mathbf{S}_{i} = \mathbf{S}(\mathbf{r}_i)$ localized on lattice sites $\mathbf{r}_i$. The magnetic Hamiltonian consists of exchange (Heisenberg) interactions, uniaxial anisotropy and a Zeeman interaction due to an external magnetic field $\mathbf{H}=H \mathbf{z}$. It reads as~[\onlinecite{Rezende}] 
\begin{align}
\mathcal{H}=  2 J \sum_{i \neq j}  \mathbf{S}_{i} \cdot \mathbf{S}_{j} - K \sum_{i} (S^{z}_{i})^{2} -  \sum_{i} \mathbf{S}_{i} \cdot \mathbf{H}\,,
\label{200}
\end{align}
where $J$ and $K>0$ parametrize, respectively, the exchange interaction and anisotropy strength.

Denoting the spins of the up and down sublattices by subscripts A and B, we can introduce the Holstein-Primakoff transformation  in the linear approximation as
\begin{align}
S^{+}_{A i}&=(2S)^{1/2} a_{i}, \; \; \; S^{-}_{A i}=(2S)^{1/2} a^{\dagger}_{i},  \;\; \; S^{z}_{A i}= S-a^{\dagger}_{i} a_{i}\,, \nonumber \\
S^{+}_{B i}&=(2S)^{1/2} b^{
\dagger}_{i}, \; \; \; S^{-}_{B i}=(2S)^{1/2} b_{i},  \;\; \; S^{z}_{B i}= - S-b^{\dagger}_{i} b_{i}\,,
\label{HP}
\end{align}
where $a_{i} (a^{\dagger}_{i})$ and $b_{i} (b^{\dagger}_{i})$ annihilate (create) a magnon at the lattice site $\mathbf{r}_{i}$ and obey bosonic commutation relations. Making use of the Fourier transformation
\begin{align}
a_{i}=N^{-1/2} \sum_{k} e^{i \mathbf{k} \cdot \mathbf{r}_{i}} a_{k}, \; \; \; b_{i}=N^{-1/2} \sum_{k} e^{i \mathbf{k} \cdot \mathbf{r}_{i}} b_{k}\,,
\end{align}
and retaining only quadratic terms in the bosonic operators, the resulting Hamiltonian can be diagonalized by the following Bogoliubov transformation
\begin{align}
a_{k}=&u_{k} \alpha_{k} - v_{k} \beta^{\dagger}_{-k}\,, \; \; \; a^{\dagger}_{k}= u_{k} \alpha^{
\dagger}_{k} - v_{k} \beta_{-k}\,, \nonumber \\
b_{k}=& u_{k} \beta_{k} - v_{k} \alpha^{\dagger}_{-k}\,, \; \; \;  b^{
\dagger}_{k}= u_{k} \beta^{\dagger}_{k} - v_{k} \alpha_{-k}\,,
\label{Bogoliubov}
\end{align}
with parameters
\begin{align}
u_{k}= \left( \frac{\omega_{ZB}+ \omega_{k}}{2 \omega_{k}} \right)^{\frac{1}{2}}, \; \; \; v_{k}= \left( \frac{\omega_{ZB} - \omega_{k}}{2 \omega_{k}} \right)^{\frac{1}{2}}\,,
\end{align}
where $\omega_{ZB}= H_{E}+H_{A}$, with   $H_{E}=2J z/\gamma \hbar$ and $H_{A}=(2S-1)K/ \gamma \hbar$ being, respectively, the exchange and anisotropy fields, and 
\begin{align}
\omega_{k}=\gamma \left[ \Delta^2 + H^2_{E} (1-\gamma^2_{k}) \right]^{\frac{1}{2}}\,, \; \;  \Delta=\left[ H_{A}(2H_{E}+H_{A})\right]^{\frac{1}{2}}\,.
\end{align}
Here, we have introduced   $\gamma_{k}=(1/z) \sum_{\delta} e^{i \mathbf{k} \cdot \boldsymbol{\delta}}$, where  $z$ is the number of the nearest neighbors spins and $\boldsymbol{\delta}$ is the vector connecting them.  
%With nearest-neighbor interactions in a simple cubic crystal
%having a lattice parameter $a$, one has
%\begin{align}
%\gamma_{\mathbf{k}}&=\frac{1}{3} \left( \cos k_{x} a + \cos k_{y} a +\cos k_{z} a \right)\, \nonumber \\
%&\stackrel{k a \ll 1}{\simeq} 1-\frac{(ka)^2}{6}\,.
%\label{142}
%\end{align}
In terms of the Bogoliubov quasi-particles~(\ref{Bogoliubov}), the Hamiltonian~(\ref{200}) simplifies to
\begin{align}
\mathcal{H}=\sum_{k} \hbar ( \omega_{k, \alpha} \alpha^{
\dagger}_{k} \alpha_{k} + \omega_{k, \beta} \beta^{
\dagger}_{k} \beta_{k})\,,
\label{245}
\end{align}
with $ \omega_{k, \alpha (\beta)}= \omega_{k} \pm \gamma H$ is the dispersion of the $\alpha$ ($\beta$) Bogoliubov quasi-particles. By rewriting the $z$-component of the spin angular momentum $S^{z}=\sum_{i} \left( S^{z}_{iA} + S^{z}_{iB} \right)$ in terms of the Bogoliubov quasi-particles~(\ref{Bogoliubov}) as
\begin{align}
S^{z}=\sum_{k} \hbar \left( -\alpha^{\dagger}_{k} \alpha_{k} + \beta^{\dagger}_{k} \beta_{k}\right)\,,
\label{Sz}
\end{align}
we can identify the bosonic operator $\alpha^{\dagger}$($\beta^{\dagger}$) as the one creating a magnon with spin angular momentum $\mp \hbar$. Based on our result~(\ref{91}), the Bose-Einstein distribution function of the $\alpha$ ($\beta$) antiferromagnetic mode can written as
\begin{align}
\langle \alpha^{\dagger}_{k} \alpha_{k}  \rangle &= n_{\text{BE}} \left( \frac{\hbar \omega_{k, \alpha}+\mu}{k_{B}T}  \right)= \frac{1}{e^{\beta (\hbar \omega_{k,\alpha} + \mu)}-1}\,, \nonumber \\
\langle \beta^{\dagger}_{k} \beta_{k}  \rangle &= n_{\text{BE}} \left( \frac{\hbar \omega_{k, \beta}-\mu}{k_{B}T} \right)= \frac{1}{e^{\beta (\hbar \omega_{k, \beta} - \mu)}-1}\,,
\label{142}
\end{align}
$\langle ... \rangle$ stands for the equilibrium (thermal) average. Equation~(\ref{142}), which constitutes a central result of this work, is valid whereas the exchange coupling $\propto J$ that drives the thermalization of the magnon gas is much stronger than the non-spin-conserving interactions. 

Previous work has  addressed the statistics of the eigenmodes of the Hamiltonian~(\ref{200}), heuristically assigning two distinct chemical potentials to them~[\onlinecite{Erik}]. The latter assumption implies that the interactions thermalizing each mode separately are much stronger than the ones driving the whole system to thermal equilibrium. However, Eqs.~(\ref{200}) and (\ref{Bogoliubov}) show that the magnon-magnon scattering processes  driving the thermalization of each magnon species and of the whole system are both rooted in the exchange coupling. Moreover,  $U$(1) symmetry implies the conservation of the $z$-component, $S_{z}$, of the total spin. As shown by Eq.~(\ref{Sz}), this corresponds to the conservation of the population imbalance between the two magnon species, which is associated to a single chemical potential. 

%Electrically Driven Bose-Einstein Condensation of Magnons in Antiferromagnets (cite). Other paper of Scott?

\subsection{Antiferromagnetic resonance}
In the long wavelength and low temperature limit, i.e., the continuum limit, the fields that describe the collective magnetic dynamics of the Hamiltonian~(\ref{200}) are the spin densities associated with the two sublattices, i.e., $\mathbf{s}_{A,B}(\mathbf{r},t)=s \mathbf{n}_{A,B}(\mathbf{r},t)$, where $s=S/V$ is the saturation spin density and $\mathbf{n}_{A (B)}(\mathbf{r},t)$ a directional unit vector. Throughout this section, we focus explicitly on the spin dynamics associated with the lowest-energy mode, i.e., $k=0$. 
The corresponding phenomenological Landau-Lifshitz equations read as~[\onlinecite{keffer}]
\begin{align}
\frac{d \mathbf{n}_{A (B)}}{dt}=&-\gamma \mathbf{n}_{A (B)} \times \big[\mathbf{H} \pm \mathbf{H}_{A}   -  H_{E} \mathbf{n}_{(B)A}  \big]\,,
\label{152}
\end{align}
with $\mathbf{H}_{A}=H_{A} \hat{\mathbf{z}}$.  In linear response, we can decompose each sublattice order parameter into an equilibrium and oscillating part, i.e., $\mathbf{n}_{A(B)}=\pm \hat{\mathbf{z}} + e^{i \omega t} \mathbf{n}_{\bot, A(B)}$.
We can rewrite Eq.~(\ref{152}) in terms of its eigenvectors $\mathbf{n}_{\alpha, \beta}$ as
\begin{align}
\frac{d \mathbf{n}_{\alpha (\beta)}}{dt}= - \gamma \mathbf{n}_{\alpha (\beta)} \times \mathbf{H}_{\alpha (\beta)}\,,
\label{162}
\end{align}
where $\gamma \mathbf{H}_{\alpha (\beta)}= \pm \hat{\mathbf{z} } \omega_{\alpha (\beta)} $, with $ \omega_{\alpha (\beta)} \equiv \omega_{k=0, \alpha (\beta)}$.  From Eq.~(\ref{162}), one can see that the order parameter $\mathbf{n}_{\alpha (\beta)}$ precesses counterclockwise (clockwise) around the $\hat{\mathbf{z}}$ axis with frequency $\omega_{\alpha (\beta)}$.
We  account for dissipation mechanisms by introducing the dimensionless Gilbert damping parameters $\alpha_{\alpha, \beta}\ll 1 $. Including an AC magnetic field  $\mathbf{h}(t)$ transverse to the equilibrium orientation of the spin densities,  Eq.~(\ref{162}) becomes
\begin{align}
\frac{d \mathbf{n}_{\alpha(\beta)}}{dt}= &- \gamma \mathbf{n}_{\alpha(\beta)} \times \left[ \mathbf{H}_{\alpha(\beta)} + \mathbf{h}(t) \right]\nonumber \\
&-\alpha_{\alpha(\beta)} \mathbf{n}_{\alpha(\beta)}  \times \frac{d \mathbf{n}_{\alpha(\beta)}}{dt}\,.
\label{17}
\end{align}
We introduce a right(left)-hand ($\pm$)  circularly-polarized AC magnetic field as
\begin{align}
\mathbf{h}_{\pm}(t)=h (\cos \omega t, \pm \sin \omega t)\,,
\label{186}
\end{align}
where $h$ is the field amplitude, with $\gamma h \ll \omega_{\alpha (\beta)}$. Plugging Eq.~(\ref{186}) into Eq.~(\ref{17}) and using the method of the complex amplitudes~[\onlinecite{gurevich}], we find that the right(left)-hand circularly polarized  field $\mathbf{h}_{\pm}(t)$ can excite resonantly the order parameter $\mathbf{n}_{\alpha (\beta)}(t)$, i.e., 
\begin{align}
n_{\alpha (\beta) x}(t)&=\frac{(\omega_{\alpha (\beta)} -\omega) \cos \omega t  -\alpha_{\alpha(\beta)}  \omega \sin \omega t }{(\omega_{\alpha (\beta)}-\omega)^2+(\alpha_{\alpha(\beta)} \omega)^2} \gamma h \,, \nonumber \\
n_{\alpha (\beta) y}(t)&= \frac{(\omega_{\alpha (\beta)} -\omega) \sin \omega t \pm \alpha_{\alpha(\beta)}  \omega \cos \omega t}{(\omega-\omega_{\alpha (\beta)})^2+(\alpha_{\alpha(\beta)}  \omega)^2} \gamma h\,,
\label{192}
\end{align}
while driving the dynamics of the order parameter $\mathbf{n}_{\alpha (\beta)}(t)$ with a left(right)-hand circularly polarized  field $\mathbf{h}_{\pm}(t)$ leads to a non-resonant dynamics, i.e., 
\begin{align}
n_{\alpha (\beta) x}(t)&=\frac{(\omega_{\alpha (\beta)} +\omega) \cos \omega t  - \alpha_{\alpha(\beta)}  \omega \sin \omega t }{(\omega_{\alpha (\beta)}+\omega)^2+(\alpha_{\alpha(\beta)}  \omega)^2} \gamma h \,, \nonumber \\
n_{\alpha (\beta) y}(t)&= \frac{\mp (\omega_{\alpha (\beta)} +\omega) \sin \omega t \mp \alpha_{\alpha(\beta)}  \omega \cos \omega t}{(\omega+\omega_{\alpha (\beta)})^2+(\alpha_{\alpha(\beta)}  \omega)^2} \gamma h\,.
\label{194}
\end{align}

\section{Thermomagnonic torques}

In this section, we construct a general phenomenology describing the interplay between the coherent and incoherent spin dynamics in a bipartite axially-symmetric antiferromagnet. At zero temperature, the total spin density is given by $\mathbf{s}=s (\mathbf{n}_{\alpha} +\mathbf{n}_{\beta})$. Thermal fluctuations engender a finite thermal magnon density of both antiferromagnetic  modes, i.e., $\tilde{n}_{\alpha}$ and $\tilde{n}_{\beta}$, which carry, respectively, $- \hbar$ and $+\hbar$ angular momentum.  Thus, the effective spin density at finite temperatures can be written as $\tilde{\mathbf{s}}=\tilde{s}_{\alpha} \mathbf{n}_{\alpha} + \tilde{s}_{\beta} \mathbf{n}_{\beta}$, with $\tilde{s}_{\alpha (\beta)}=s-\tilde{n}_{\alpha (\beta)}$.

If the coherent texture is smooth on the scale of the thermal magnon wavelength, the hydrodynamic variables that describe the system are the order parameters $\mathbf{n}_{\alpha, \beta}$ and the imbalance between the thermal-magnon densities, i.e.,  $\tilde{n}=\tilde{n}_{\beta}-\tilde{n}_{\alpha}$. 
Hence, the free energy  of the antiferromagnet can be written as $\mathcal{F} \equiv\mathcal{F}[\mathbf{n}_{\alpha}, \mathbf{n}_{\beta}, \tilde{n}]$. We can identify the effective transverse field (at finite wave vector) $\mathbf{H}_{\alpha ( \beta)} \equiv \delta_{\mathbf{n}_{\alpha (\beta)}}\mathcal{F}[\mathbf{n}_{\alpha}, \mathbf{n}_{\beta}, \tilde{n}]$ and the magnon chemical potential $\mu \equiv \delta_{\tilde{n}} \mathcal{F}[\mathbf{n}_{\alpha}, \mathbf{n}_{\beta}, \tilde{n}]$ as the thermodynamics forces  conjugate to the order parameters $\mathbf{n}_{\alpha, \beta}$ and the thermal magnon imbalance $\tilde{n}$, respectively.
The phenomenological equations of motion can be written  in terms of a $5 \times 5$ linear-response matrix $\mathbf{L}$ as

\begin{align}
\begin{pmatrix} \dot{\mathbf{n}}_{\alpha} \\ \dot{\mathbf{n}}_{\beta} \\\tilde{n}\end{pmatrix}=  \begin{pmatrix} \mathbf{L}_{\alpha \alpha} & \mathbf{L}_{\alpha \beta} & \mathbf{L}_{\alpha \tilde{n}} \\  \mathbf{L}_{\beta \alpha} & \mathbf{L}_{\beta \beta} & \mathbf{L}_ {\beta \tilde{n}} \\ \mathbf{L}_{\tilde{n} \alpha} &  \mathbf{L}_{\tilde{n} \beta} &  \mathbf{L}_{\tilde{n} \tilde{n}} \\ \end{pmatrix} \begin{pmatrix} \mathbf{H}_{\alpha} \\ \mathbf{H}_{\beta} \\ \mu \end{pmatrix}\,,
\end{align}
where $\mathbf{L}_{\alpha (\beta) \tilde{n}}$ is a $2 \times 2$ matrix that describes the torques exerted by the thermal magnon imbalance on the coherent order parameter $\mathbf{n}_{\alpha (\beta)}$. The reciprocal counterpart $\mathbf{L}_{\tilde{n} \alpha (\beta)}$ accounts for the spin pumping by the order parameter $\mathbf{n}_{\alpha, \beta}$ into the thermal magnon cloud.  Following the approach of Ref.~[\onlinecite{Flebus2016}], we determine the explicit form of these matrices invoking  Onsager reciprocity, which dictates
\begin{align}
[ \mathbf{L}_{\alpha(\beta) \tilde{n}}(\mathbf{n}_{\alpha (\beta)})  ]_{ij}=[ \mathbf{L}_{\tilde{n} \alpha(\beta) }(-\mathbf{n}_{\alpha (\beta)})  ]_{ji}\,,
\end{align}
and the Neumann principle, which requires the equations of motion to be invariant for a rotation of the magnetic variables around the $\hat{\textbf{z}}$ axis. Furthermore, $U$(1) symmetry implies the conservation of the $z$-component of the spin density, i.e., 
\begin{align}
\tilde{s}_{\alpha} \dot{n}_{\alpha,z} + \tilde{s}_{\beta} \dot{n}_{\beta,z}+\dot{\tilde{n}}=0\,.
\end{align}
Focusing only on the  spin transfer and pumping processes between the incoherent and coherent spin dynamics, we can rewrite
\begin{align}
\hbar \dot{\mathbf{n}}_{\alpha}=&- \eta_{\alpha} \mathbf{n}_{\alpha} \times ( \hbar \dot{\mathbf{n}}_{\alpha}-\mu \mathbf{z} \times \mathbf{n}_{\alpha} ) \,, \\
\hbar \dot{\mathbf{n}}_{\beta}=&- \eta_{\beta} \mathbf{n}_{\beta} \times ( \hbar \dot{\mathbf{n}}_{\beta}-\mu \mathbf{z} \times \mathbf{n}_{\beta} ) \,, \\ 
\dot{\tilde{n}}
=&-\eta_{\alpha} \tilde{s}_{\alpha} \mathbf{z} \cdot \mathbf{n}_{\alpha} \times ( \dot{\mathbf{n}}_{\alpha}- \frac{\mu}{\hbar} \mathbf{n}_{\alpha}\times \mathbf{z}) \nonumber \\
& -\eta_{\beta} \tilde{s}_{\beta} \mathbf{z} \cdot \mathbf{n}_{\beta} \times ( \dot{\mathbf{n}}_{\beta}- \frac{\mu}{\hbar} \mathbf{n}_{\beta}\times \mathbf{z})\,. \label{193}
\end{align}
Here, $\eta_{\alpha(\beta)} \ll 1$ is a phenomenological (dimensionless) coefficient parametrizing the strength of interactions between the order parameter $\mathbf{n}_{\alpha (\beta)}$ and the thermal magnon cloud.  The Onsager reciprocal relations between the components of the transverse magnetization dynamics, i.e., 
\begin{align}
\left[ \mathbf{L}^{\mathbf{n}_{\alpha (\beta)} \mathbf{n}_{\alpha (\beta)}}(\mathbf{n}_{\alpha (\beta)}) \right]_{ij}=\left[ \mathbf{L}^{\mathbf{n}_{\alpha (\beta)} \mathbf{n}_{\alpha (\beta)}}(\mathbf{n}_{\alpha (\beta)}) \right]_{ji}\,,
\end{align}
constrain the parameter $\eta_{\alpha (\beta)}$ to be an even function of $n_{\alpha (\beta),z}$.

\section{Local Measurement of the  chemical potential}

In this section, we unveil the relation between the relaxation rate of a NV center set nearby a $U$(1)-symmetric bipartite antiferromagnet and the chemical potential of the antiferromagnetic system. Then, making use of the results derived in Secs.~II C and Sec.~III, we show that the chemical potential can be controlled by means of a AC field driving resonantly one of the two antiferromagnetic modes.

\subsection{NV-center relaxometry of an antiferromagnet}

\begin{figure}[t!]
\includegraphics[width=1\linewidth]{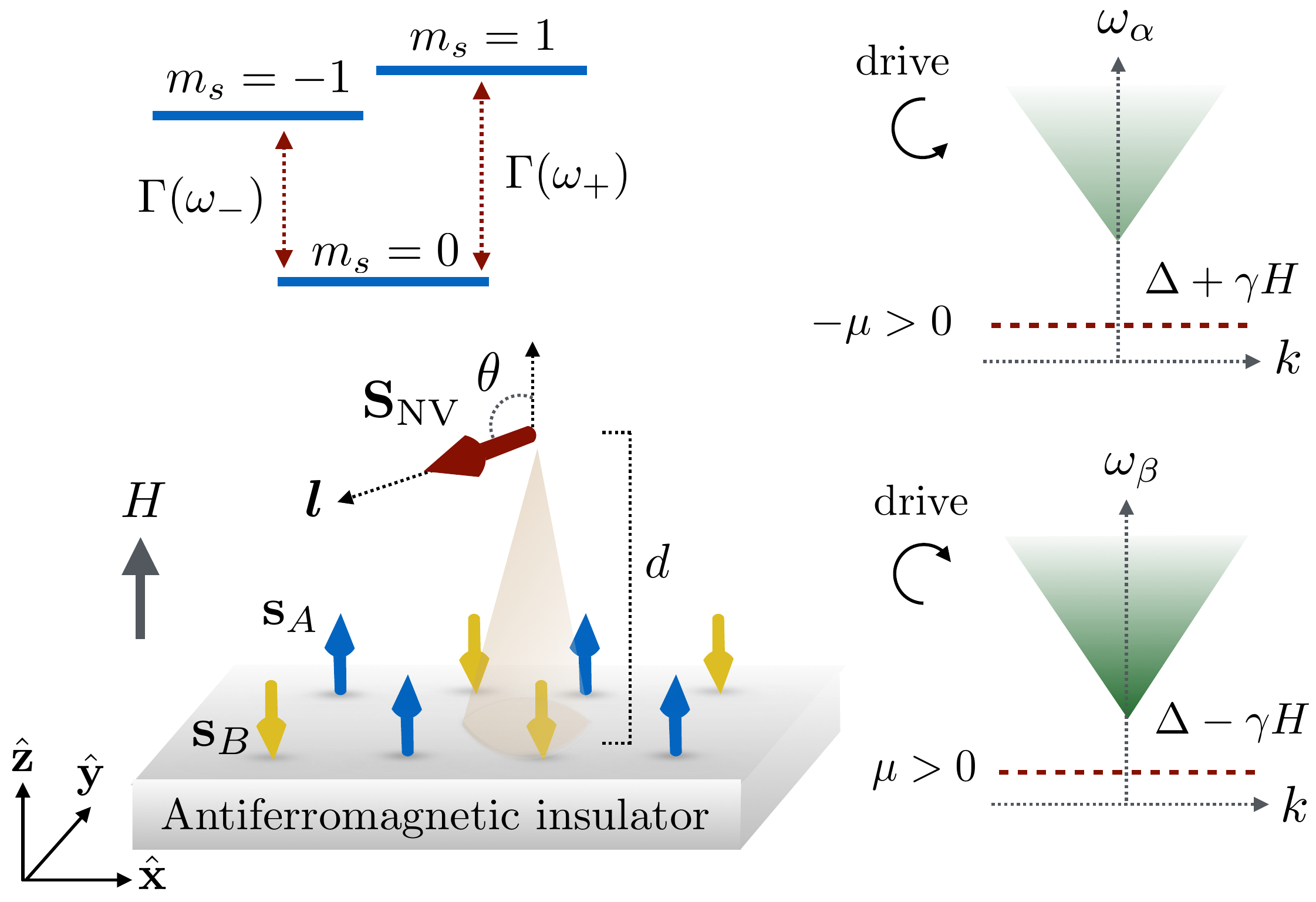}
\caption{Set up for controlling and probing the chemical potential of an antiferromagnetic insulator. Bottom figure: A NV-center spin $\mathbf{S}_{\text{NV}}$ lies at a height $d$ above an antiferromagnetic insulator and it is oriented along its anisotropy axis $\boldsymbol{l}$, with  $\boldsymbol{l} \cdot \hat{\mathbf{z}}=\cos \theta$. A static magnetic field $H$ is applied along the $z$ direction. The interactions between the QI spin and the sublattice  spin densities $\mathbf{s}_{A}$ and $\mathbf{s}_{B}$ induce QI transitions between the spin states $m_{s}=0 \leftrightarrow \pm 1$ with energy loss or gain of  $\omega_{\pm}$ at the rate $\Gamma(\omega_{\pm})$. Top figure: a right(left)-handed circularly polarized field can drive resonantly the coherent spin dynamics of the  antiferromagnetic mode $\alpha$ ($\beta$), which, in turn, gives rise to a finite negative (positive) chemical potential. If one of the NV-center resonance frequencies  is equal or larger than the magnetic gap of the $\alpha$ ($\beta$) mode, i.e., $\omega_{\pm} \geq \Delta \pm \gamma H$,  the relaxation rate $\Gamma (\pm \omega)$ provides a direct measurement of the chemical potential. }
\label{Fig2}
\end{figure}

In the setup we envision, illustrated in Fig.~(\ref{Fig2}), the NV center lies at a height $d$ above the antiferromagnetic sample and it is oriented along its own anisotropy axis $\boldsymbol{l}$, with $  \boldsymbol{l} \cdot \mathbf{z}=\cos \theta$. The NV center can be modelled as a single-spin system with spin $|\mathbf{S}_{\text{NV}}|=1$ (hence, with secondary spin quantum number $m_{s}=0, \pm 1$) ~[\onlinecite{Casola}]. The latter couples, via Zeeman interaction, to  the magnetostatic field generated by the fluctuations of the antiferromagnetic spin density $\mathbf{s}(\mathbf{r})$, which can be written as
\begin{align}
\mathbf{B}(\mathbf{r}_{\text{NV}})=\gamma_{\text{NV}} \int d^2 \mathbf{r} \; \mathcal{D}(\mathbf{r}, \mathbf{r}_{\text{NV}}) \mathbf{s}(\mathbf{r})\,,
\label{208}
\end{align}
where $\gamma_{\text{NV}}$ and $\mathbf{r}_{\text{NV}}=(0,0,d)$ are, respectively, the gyromagnetic ratio and the position of the NV center, while  $\mathcal{D}$ is the tensorial magnetostatic Green's function~[\onlinecite{slavin}]. Up to leading order in perturbation theory, the coupling between the NV-center spin and the magnetostatic field~(\ref{208}) gives rise to the transitions $m_{s}=0 \leftrightarrow \pm 1$ at the frequencies $\omega_{\pm}=\Delta_{\text{NV}} \pm \gamma H \cos\theta$, where $\Delta_{\text{NV}}$ is the NV-center zero-field splitting. The corresponding relaxation rate is given by~[\onlinecite{FlebusPRL2018}]
\begin{align}
\Gamma (\omega_{\pm})=&f(\theta) \int^{\infty}_{0} dk \;  k^{3} e^{-2kd}  \big[ 4 C_{zz}(k, \omega_{\pm}) \nonumber \\ + & C_{-+}(k, \omega_{\pm})+ C_{+-}(k, \omega_{\pm})\big]\,,
 \label{relaxationrate}
\end{align}
with $f(\theta)=  (\gamma \tilde{\gamma})^2  ( 5 -\cos2\theta)/64 \pi$. Here, $C_{m n}(k, \omega_{\pm})$ is the real part of the Fourier transform of the spin-spin correlator  
\begin{align}
C_{m n}
(\mathbf{r}_{i},\mathbf{r}_{j}; t)=\langle \{ s^{m}(\mathbf{r}_{i},t), s^{n}(\mathbf{r}_{j},0) \} \rangle\,.
\label{257}
\end{align}
For $m,n=\pm (z)$, the spin-spin correlator~(\ref{257}) describes magnetic noise transverse (longitudinal) to the magnetic symmetry axis $\hat{\mathbf{z}}$. From Eqs.~(\ref{HP}) and (\ref{Bogoliubov}), one can see that the transverse (longitudinal) spin-spin correlators account for one (two)-magnon processes, i.e., $C_{+-}(k) \propto \alpha^{\dagger}_{k}, \beta^{\dagger}_{k}$ ($C_{zz}(k) \propto \alpha_{k}^{\dagger} \alpha_{k}, \beta_{k}^{\dagger} \beta_{k}$).
We set the NV frequency in the region corresponding to a non-vanishing spectral density of at least one antiferromagnetic mode, i.e., $\omega_{\pm} \geq \Delta - \gamma H$. The contribution of one-magnon excitations to the relaxation rate dominates and we can neglect, in first instance, multi-magnon processes~[\onlinecite{Du2017},\onlinecite{FlebusPRL2018}].  In this limit, Eq.~(\ref{relaxationrate}) becomes
\begin{align}
\Gamma (\omega_{\pm})&=4s  f(\theta) \int^{\infty}_{0} dk \;  k^{3} e^{-2kd}  (u_{k} - v_{k})^2 \; \nonumber \\
&\times 
\left[  n_{\text{BE}}\left( \frac{ \hbar \omega_{k, \alpha} - \mu}{k_{B}T}  \right)   + \frac{1}{2} \right] \delta(\omega_{k,\alpha} -\omega_{\pm}) \nonumber \\
&+4s  f(\theta) \int^{\infty}_{0} dk \;  k^{3} e^{-2kd}  (u_{k} - v_{k})^2 \; \nonumber \\
&\times 
\left[  n_{\text{BE}}\left( \frac{ \hbar \omega_{k, \beta} + \mu}{k_{B}T}  \right)   + \frac{1}{2} \right] \delta(\omega_{k,\beta} -\omega_{\pm}) \,.
\label{266}
\end{align}
Equation~(\ref{266}) shows that, when the resonance frequency of the NV center equals the dispersion of one antiferromagnetic mode,  the NV-center relaxation rate is directly proportional to the distribution function of the latter, and, thus,  provides a measurement of the chemical potential.

\subsection{Control of the chemical potential}

A diffusion equation for the $z$-component of the spin density can be written as~[\onlinecite{twofluid}]
\begin{align}
\delta \dot{\tilde{n}} + \boldsymbol{\nabla} \cdot \mathbf{j}= - g \mu\,,
\label{229}
\end{align}
where $\mathbf{j}=-\sigma \boldsymbol{\nabla} \mu$ is the spin current, $\sigma$ the bulk spin conductivity and $g$ parametrizes the relaxation of the spin density by the phononic environment. Including the interplay between the coherent and incoherent spin dynamics~(\ref{193}), Eq.~(\ref{229}) becomes
\begin{align}
\delta \dot{\tilde{n}} + \boldsymbol{\nabla} \cdot \mathbf{j}=& - g \mu -\eta_{\alpha} \tilde{s}_{\alpha} \mathbf{z} \cdot \mathbf{n}_{\alpha} \times ( \dot{\mathbf{n}}_{\alpha}- \frac{\mu}{\hbar} \mathbf{n}_{\alpha}\times \mathbf{z}) \nonumber \\
& -\eta_{\beta} \tilde{s}_{\beta} \mathbf{z} \cdot \mathbf{n}_{\beta} \times ( \dot{\mathbf{n}}_{\beta}- \frac{\mu}{\hbar} \mathbf{n}_{\beta}\times \mathbf{z})\,.
\label{232}
\end{align}
In the setup we envision, the chemical potential can be controlled via an AC magnetic field, which we treat as homogeneous throughout the sample. Thus, we can set $\boldsymbol{\nabla} \mu=0$.
Assuming the magnon-phonon coupling to be comparable with the interactions between the coherent and incoherent spin dynamics~[\onlinecite{Du2017}], i.e., $\eta_{\alpha, \beta} \sim g$,   we  neglect  the corrections to the chemical potential due to the term $\propto \eta_{\alpha (\beta)} \mu$ . Thus, for a steady state, we can rewrite   Eq.~(\ref{232}) as
\begin{align}
\mu= -  \frac{ \eta_{\alpha} \tilde{s}_{\alpha}}{g} \langle  \mathbf{z} \cdot \mathbf{n}_{\alpha} \times \dot{\mathbf{n}}_{\alpha} \rangle - \frac{  \eta_{\beta} \tilde{s}_{\beta}}{g} \langle   \mathbf{z} \cdot \mathbf{n}_{\beta} \times \dot{\mathbf{n}}_{\beta} \rangle\,,
\label{282}
\end{align}
where $\langle ... \rangle$  denotes the time average over a full precession cycle. As shown by Eq.~(\ref{192}), a right(left)-hand circularly polarized AC field can resonantly drive the dynamics of the parameter $\mathbf{n}_{\alpha (\beta)}$.  When the dynamics of the order parameter $\mathbf{n}_{\alpha (\beta)}$ is driven resonantly, the contribution to the chemical potential~(\ref{282}) due to the term $\propto \langle \mathbf{z} \cdot \mathbf{n}_{\beta (\alpha)} \times \dot{\mathbf{n}}_{\beta (\alpha)}  \rangle$ is negligible. Thus, when the magnetization dynamics is driven by a right(left)-hand circularly polarized field set at frequency $\omega=\omega_{\alpha (\beta)}$, Eq.~(\ref{282}) reads as
\begin{align}
\mu=\mp  \frac{\tilde{s}_{\alpha (\beta)} \eta_{\alpha (\beta)} (\gamma h)^2}{g \alpha_{\alpha(\beta)} \omega_{\alpha (\beta)}}  \,.
\label{287} 
\end{align}
As illustrated in Fig.~(\ref{Fig2}), Eq.~(\ref{287}) shows that driving the AFMR of the magnon mode that carries spin angular momentum $\pm \hbar$ engenders a finite positive (negative) chemical potential. The latter can be probed experimentally via the NV-center relaxation rate~(\ref{266}). \\
\section{Conclusion and outlook}

In this work, we established a rigorous definition of the magnon chemical potential of a $U$(1)-symmetric antiferromagnetic insulating system. Specifically, we found that the two antiferromagnetic modes are described by an equal and opposite chemical potential. 
We developed  a general phenomenology that accounts for the coupled dynamics of the antiferromagnetic order parameters and the thermal magnon cloud. Our results show that the torque exerted by the thermal magnon cloud on the order parameters and the back-action of the latters resemble the spin transfer torque and spin pumping processes at a magnetic insulator$|$normal metal interface~[\onlinecite{Berger1996},\onlinecite{Bauer2002}]. In particular, the dynamics of each magnetic order parameter can pump spin angular momentum  into the thermal magnon cloud, leading to a finite magnon chemical potential, whose sign depends on the handedness  of the order-parameter precession.
Thus, by driving resonantly the dynamics of one of the antiferromagnetic order parameters via an AC magnetic field,  one can tune the chemical potential of the system. In Ref.~[\onlinecite{Du2017}], the authors found that the magnon chemical potential of YIG can reach, as function of the AC drive amplitude, its maximum value, i.e., corresponding to magnon BEC. This result suggests that, for antiferromagnetic materials with comparable magnetic gaps,  magnon BEC could be experimentally realized in an analogous way. In an antiferromagnetic system,  the magnon species undergoing BEC can be selected by the handedness of the drive.  Alternatively, selective magnon BEC could be achieved by means of a spin accumulation in an adjacent metal. By reversing the direction of the charge current flowing into the metal, one can tune the orientation of the spin accumulation induced via the Spin Hall effect~[\onlinecite{SHE}]. A spin accumulation set along the $\pm z$ direction will exert a spin transfer torque on the antiferromagnetic mode with positive (negative) chemical potential, possibly triggering its condensation~[\onlinecite{bender}].  The hydrodynamics and transport properties of this partially condensed state are an interesting direction for future research.

Furthermore, we discuss NV center relaxometry as a way to probe the antiferromagnetic chemical potential. The proposed experimental set-up requires the NV-center frequency to be comparable or higher than the antiferromagnetic gap. Since the maximum NV-center frequency is of the order of $\sim 100$ GHz~[\onlinecite{Casola}], two relevant materials for the proposal presented here  are the antiferromagnetic insulators $\text{RbMnF}_{3}$ and $\text{KNiF}_3$. These simple cubic Heisenberg antiferromagnets display a relatively low magnetic gap due to an anisotropy magnetic field of, respectively, $H_{A}=4.5$  Oe and $H_{A}=80$ Oe, five orders of magnitude smaller than the corresponding exchange field $H_{E}$~[\onlinecite{miedema}, \onlinecite{yamaguchi}]. 

Finally, throughout this work, we have assumed that the static magnetic field is far below the critical field value that triggers a reorientation transition in easy-axis antiferromagnets, i.e., below the spin-flop transition. Future work should address the statistics of the magnetic excitations throughout this transition.

\section{Acknowledgements} 
The author thanks C. Du,  Y. Tserkovnyak and R. A. Duine for insightful discussions. B. F. was supported by the Dutch Science Foundation (NWO) through a Rubicon grant.


\begin{thebibliography}{99}

\bibitem{review} M. Wu and A. Hoffmann, \textit{Recent advances in magnetic insulators - From spintronics to microwave applications}, Solid State Physics \textbf{64} (Academic Press, 2013).

\bibitem{Cornelissen2015}  L. J. Cornelissen, J. Liu, R. A. Duine, J. Ben Youssef, and B. J.
van Wees, Nat. Phys. \textbf{11}, 1022 (2015).
 
\bibitem{Giles2015} B. L. Giles, Z. Yang, J. S. Jamison, and R. C. Myers,
Phys. Rev. B \textbf{92}, 224415 (2015). 
 
\bibitem{Gross2015} S. T. B. Goennenwein, R. Schlitz, M. Pernpeintner, K. Ganzhorn,
M. Althammer, R. Gross, and H. Huebl, Appl. Phys. Lett. \textbf{107},
172405 (2015). 

\bibitem{Cornelissen2016} L. J. Cornelissen, K. J. H. Peters, G. E. W. Bauer, R. A. Duine, and B. J. van Wees, Phys. Rev. B \textbf{94}, 014412 (2016).
 
\bibitem{Du2017} C. Du, T. Van der Sar, T. X. Zhou, P. Upadhyaya, F.
Casola, H. Zhang, M. C. Onbasli, C. A. Ross, R. L. Walsworth, Y. Tserkovnyak, and A. Yacoby, Science \textbf{357}, 195 (2017). 
 
\bibitem{Flebus2016} B. Flebus, P. Upadhyaya, R. A. Duine, and Y. Tserkovnyak, Phys. Rev. B \textbf{94}, 214428 (2016).
 
\bibitem{Casola} F. Casola, T. van der Sar, and A. Yacoby, Nat. Rev. Mat. \textbf{3}, 17088 (2018). 
 
\bibitem{Demokritov} 
S. O. Demokritov, V. E. Demidov, O. Dzyapko, G. A. Melkov, A. A. Serga, B. Hillebrands, and A. N. Slavin,
Nature \textbf{443}, 430 (2006).  
 
\bibitem{Sonin} E. B. Sonin,  Adv. Phys. \textbf{59}, 181 (2010).
 
\bibitem{Takei} S. Takei, and Y. Tserkovnyak,  Phys. Rev. Lett. \textbf{112}, 227201 (2014).
 
\bibitem{Yuan} W. Yuan, Q. Zhu, T. Su, Y. Yao, W. Xing, Y. Chen, Y. Ma, X. Lin, J. Shi, R. Shindou,
X. C. Xie, and W. Han, Sci. Adv. \textbf{4}, 1098 (2018). 

\bibitem{Klaui} R. Lebrun, A. Ross, S. A. Bender, A. Qaiumzadeh, L. Baldrati, J. Cramer, A. Brataas, R. A. Duine, and M. Klaui,  Nature \textbf{561}, 222 (2018). 

\bibitem{Baltz} V. Baltz, A. Manchon, M. Tsoi, T. Moriyama, T. Ono, and Y. Tserkovnyak,
Rev. Mod. Phys. \textbf{90}, 015005 (2018).

\bibitem{Berger1996} J. C. Slonczewski, J. Magn. Magn. Mater. \textbf{159}, L1 (1996); L. Berger, Phys. Rev. B \textbf{54}, 9353 (1996).
\bibitem{Bauer2002} Y. Tserkovnyak, A. Brataas, and G. E. W. Bauer, Phys.  Rev. Lett. \textbf{88}, 117601 (2002).

\bibitem{Rezende} S. M. Rezende, R. L. Rodr\'iguez-Su\'arez, and A. Azevedo, Phys. Rev. B \textbf{93}, 014425 (2016).


\bibitem{Erik} E. Lohaugen Fjaerbu, N. Rohling, and A. Brataas,
Phys. Rev. B \textbf{95}, 144408 (2017).

\bibitem{keffer}  F. Keffer and C. Kittel, Phys. Rev. \textbf{85}, 329 (1952).

%\bibitem{beforekamra} Here, we have implicitely assumed that the sublattice spin densities are subjected to the same dissipation. Corrections to this approximation for antiferromagnetic systems with non-identical sublattices have been extensively discussed in Ref.~[\onlinecite{kamraGilbert}].
%
%\bibitem{kamraGilbert} A. Kamra, R. E. Troncoso, W. Belzig, and A. Brataas,  Phys. Rev. B \textbf{98}, 184402 (2018).

\bibitem{gurevich} A. G. Gurevich and G. A. Melkov, \textit{Magnetization Oscillations and Waves} (CRC Press, 1996).





\bibitem{slavin} K. Y. Guslienko, and A. N. Slavin, J. Magn. Magn.
Mater. \textbf{323}, 2418 (2011).

\bibitem{FlebusPRL2018} B. Flebus and Y. Tserkovnyak, Phys. Rev. Lett. \textbf{121}, 187204 (2018).

\bibitem{twofluid} B. Flebus, S. A. Bender, Y. Tserkovnyak, and R. A. Duine,  Phys. Rev. Lett. \textbf{116}, 117201 (2016).

\bibitem{SHE} J. E. Hirsch
Phys. Rev. Lett. \textbf{83}, 1834 (1999).


\bibitem{bender} S. A. Bender, R. A. Duine, and Y. Tserkovnyak,
Phys. Rev. Lett. \textbf{108}, 246601 (2012).

\bibitem{miedema} L. de Jongh and A. Miedema, Advances in Physics \textbf{23}, 1 (1974).

\bibitem{yamaguchi} H. Yamaguchi, K. Katsumata, M. Hagiwara, M. Tokunaga, H. L.
Liu, A. Zibold, D. B. Tanner, and Y. J. Wang, Phys. Rev. B \textbf{59},
6021 (1999).

\end{thebibliography}
\end{document}